\documentclass[twocolumn,aps,superscriptaddress,showpacs]{revtex4}
\usepackage{amsmath,bm}
\usepackage{graphicx}
\begin{document}

\title{Neutron removal cross section as a measure of neutron skin}

\author{D. Q. Fang}
\thanks{E-mail address: dqfang@sinap.ac.cn}

\author{Y. G. Ma}
\author{X. Z. Cai}
\author{W. D. Tian}
\author{H. W. Wang}

\affiliation{Shanghai Institute of Applied Physics,
Chinese Academy of Sciences, P. O. Box 800-204, 
Shanghai 201800, China}


\begin{abstract}
We study the relation between neutron removal cross section 
($\sigma_{-N}$) and neutron skin thickness for finite neutron 
rich nuclei using the statistical abrasion ablation (SAA) model. 
Different sizes of neutron skin are obtained
by adjusting the diffuseness parameter of neutrons
in the Fermi distribution. It is demonstrated that there is a 
good linear correlation between $\sigma_{-N}$ and the neutron skin
thickness for neutron rich nuclei. Further analysis suggests that
the relative increase of neutron removal cross section could 
be used as a quantitative measure for the neutron skin thickness 
in neutron rich nuclei.

\pacs{21.10.Gv, 25.70.Mn, 21.60.-n}
\end{abstract}

\maketitle

Neutron rich nuclei are 
expected to have a neutron skin due to the neutron excess and
the Coulomb barrier which suppresses the proton density at the 
nuclear surface. The thickness of neutron skin is defined as the 
difference between the root-mean-square (RMS) neutron and proton
radii, i.e. $S_{n}\equiv{<r_n^2>}^{1/2}-{<r_p^2>}^{1/2}$.
The neutron skin thickness dependents on the properties 
of the equation of state (EOS) of neutron rich matter. 
More neutrons are pushed to the nuclear surface with higher 
pressure, which will form thicker neutron skin.

It has been show that $S_{n}$ is found to 
be related to a constraint on the nuclear EOS within framework 
of Skyrme Hartree-Fock (SHF) and relativistic mean-field 
(RMF) theories~\cite{brow00,furn02,yosh04}. 
Strong linear correlations between $S_{n}$ and
$L$ (the slope of symmetry energy coefficient $C_{sym}$),
the ratio $L/J$ ($J$ is the symmetry energy coefficient at
the saturation density $\rho_0$), $J-a_{sym}$
($a_{sym}$ is the symmetry energy coefficient of finite nuclei)
are demonstrated~\cite{brow00,furn02,yosh04,dani03,chen05,cent09}. 
The constraint on EOS is important for improving our 
knowledge of neutron rich matter and extrapolating the EOS to 
higher neutron density, and has strong impact on the property 
of neutron stars. 
It was shown that $S_{n}$ is crucial for
the transition density from nonuniform to uniform neutron 
rich matter in the crust of neutron star~\cite{horw01}.
Reliable neutron densities are
needed as input for atomic parity violation experiments~\cite{poll99}.

Due to the significance of $S_{n}$ for studying structure 
of neutron rich nuclei, property of neutron rich 
matter and neutron star etc., measurement of $S_{n}$ has become a very 
important subject in the forefront of nuclear physics research. 
It is remarkable that 
such a measurement is so important for investigating the property of
very small entities like atomic nuclei as well as 
very big objects like neutron stars. 
As we know, the proton radius 
of the nucleus can be determined with a very high accuracy through 
the electron scattering measurement~\cite{fric95}. However, 
the experimental accuracy of neutron radius is much lower compared 
with that of the proton radius so far.

Recently, some methods have been proposed or used to measure 
the radius of neutron distribution. A model independent method 
using parity violating electron scattering to measure the neutron 
radius of $^{208}$Pb to 1\% accuracy ($\pm$0.05 fm) was proposed
at JLab~\cite{horo01}. There are methods to extract
neutron densities from proton-nucleus elastic scattering and 
antiprotonic atom data~\cite{kara02,clar03,brow07,trzc01}. 
Another method of using the relationship between the nuclear 
matter radius and the reaction cross section ($\sigma_R$) 
was shown~\cite{suzu95}. But it is also useful to have 
alternative methods of obtaining $S_{n}$ directly.
The dependence of neutron removal cross section 
($\sigma_{-N}$) on the neutron skin thickness was 
studied and strong linear correlation was observed~\cite{mach08}. 
In this work, it is demonstrated that the relative increase of 
neutron removal cross section could be used as a quantitative 
measure for $S_{n}$ by using the 
statistical abrasion ablation (SAA) model .

The SAA model was developed by Brohm {\it et al} to describe
heavy ion collisions at high energies in a picture of
quasi-free nucleon-nucleon (N-N) collisions~\cite{broh94}. 
This model takes independent N-N collisions for 
participants in an overlap zone of the two colliding nuclei 
and determines the distributions of abraded neutrons and protons. 
The spectators outside the overlapping region are 
left to move almost without being disturbed.
The probability of abrading zero nucleons is calculated and 
the resulting energy-dependent reaction cross section 
turns out to be equivalent to the successful microscopic
Glauber-type calculation. It has been demonstrated that 
the SAA model can reproduce the experimental fragment 
production cross section of heavy ion collisions very 
well~\cite{blan95,jong98} and is useful for studying the 
isospin effect and isoscaling behavior of projectile-like 
fragments~\cite{fang01,fang07} in heavy ions reactions. 
Here we give a brief description of the SAA model. 
Its details can be found in Ref.~\cite{broh94}.

In SAA model,
nuclear reaction is described as two stages which occur in two
different time scales. The first stage is
abrasion process which describes the production of 
pre-fragment with certain excitation energy ($E^*$) through 
the N-N collisions in the overlap zone of the
colliding nuclei.  The collisions are described by a picture of
interacting tubes. Assuming a binomial distribution for the
absorbed neutrons and protons of projectile in the interaction of a
specific pair of tubes, the distributions of the total abraded
neutrons and protons are determined. For an infinitesimal tube in
the projectile, the transmission probabilities for neutrons
(protons) at a given impact parameter $b$ $(\equiv|\vec{b}|)$ 
are calculated by~\cite{broh94}
\begin{equation}
t_{\mbox{k}}(\vec{r}-\vec{b})=\exp\{-[D_{\mbox{n}}^{\mbox{T}}
(\vec{r}-\vec{b})\sigma_{\mbox{nk}}+%
D_{\mbox{p}}^{\mbox{T}}(\vec{r}-\vec{b})\sigma_{\mbox{pk}}]\},
\end{equation}
where $D^{\mbox{T}}$ is thickness function of
the target, which is normalized by
$\int d^2r'D^{\mbox{T}}_{\mbox{n}}(\vec{r'})=N^{\mbox{T}}$ and
$\int d^2r'D^{\mbox{T}}_{\mbox{p}}(\vec{r'})=Z^{\mbox{T}}$ with
$N^{\mbox{T}}$ and $Z^{\mbox{T}}$ referring to the neutron and
proton number in the target respectively, the vectors $\vec{r}$, 
$\vec{r'}$ and $\vec{b}$
are defined in the plane perpendicular to the beam, and
$\sigma_{\mbox{k}'\mbox{k}}$ is the free nucleon-nucleon cross
sections (k$'$, k$=$n for neutron and k$'$, k$=$p for proton). The
thickness function of the target is given by
\begin{equation}
D^{\mbox{T}}_{\mbox{k}}(\vec{r'})=\int_{-\infty}^{+\infty}dz\rho_{\mbox{k}}
((\vec{r'}^2+z^2)^{1/2}),
\end{equation}
with $\rho_{\mbox{k}}$ being the neutron (proton) density
distribution of the target. The thickness function of the projectile
$D^{\mbox{P}}_{\mbox{k}}(\vec{r'})$ is defined in the same way. 
When the thickness function is used in the calculation, 
the variable $\vec{r'}$ equals $\vec{r}-\vec{b}$ for the target and
$\vec{r'}$ equals $\vec{r}$ for the projectile.
So the average abraded mass at a given
impact parameter $b$ is calculated by the expression
\begin{equation}
\begin{array}{ll}
\langle \Delta A(b) \rangle & = \langle \Delta N(b) \rangle + 
\langle \Delta Z(b) \rangle \\
& = \int d^2rD_{\mbox{n}}^{\mbox{P}}(\vec{r})[1-t_{\mbox{n}}(\vec{r}-\vec{b})] \\
& +\int d^2rD_{\mbox{p}}^{\mbox{P}}(\vec{r})[1-t_{\mbox{p}}(\vec{r}-\vec{b})],
\end{array}
\end{equation}
where $\langle \Delta N(b) \rangle$ and $\langle \Delta Z(b) \rangle$
are the average abraded neutron and proton number at impact parameter $b$, 
$D^{\mbox{P}}$ is thickness function of the projectile. 
The integrals extend over the plane perpendicular to the beam
direction. In a similar way, the variance of $\Delta A$ can be
determined by summing the variances of all the binomial
distributions for the tubes pairs,
\begin{equation} \label{}
    \begin{split}
    \langle [\Delta A(b)&-\langle\Delta A(b)\rangle]^2\rangle=\\
& \int d^2rD_{\mbox{n}}^{\mbox{P}}(\vec{r})
[1-t_{\mbox{n}}(\vec{r}-\vec{b})]t_{\mbox{n}}(\vec{r}-\vec{b})\\
&+ \int d^2rD_{\mbox{p}}^{\mbox{P}}(\vec{r})
[1-t_{\mbox{p}}(\vec{r}-\vec{b})]t_{\mbox{p}}(\vec{r}-\vec{b}),
    \end{split}
\end{equation}
Higher moments of the distributions of abraded neutrons and protons
could be obtained similarly. 
However, for the
present purpose, sufficient information is provided by the first and
second moments. In addition, the probability for zero nucleon loss
is important because it is related to the reaction cross section.
The deduced total transmission probability $T(b)$ and reaction cross 
section are~\cite{broh94}
\begin{equation} \label{}
    \begin{split}
    T(b)&=\exp(-\int d^2r\{D_{\mbox{n}}^{\mbox{P}}(\vec{r})D_{\mbox{n}}^{\mbox{T}}
    (\vec{r}-\vec{b})\sigma_{\mbox{nn}}\\
    &+ [D_{\mbox{n}}^{\mbox{P}}(\vec{r})D_{\mbox{p}}^{\mbox{T}}(\vec{r}-\vec{b})
    +D_{\mbox{p}}^{\mbox{P}}(\vec{r})D_{\mbox{n}}^{\mbox{T}}(\vec{r}-\vec{b})]\sigma_{\mbox{np}}\\
    &+D_{\mbox{p}}^{\mbox{P}}(\vec{r})D_{\mbox{p}}^{\mbox{T}}(\vec{r}-\vec{b})\sigma_{\mbox{pp}}
    \}),
    \end{split}
\end{equation}

\begin{equation}\label{}
    \sigma_{R}=\int{d^{2}b[1-T(b)]},
\end{equation}

\begin{figure}[b]
\includegraphics[width=7.2cm,angle=0.]{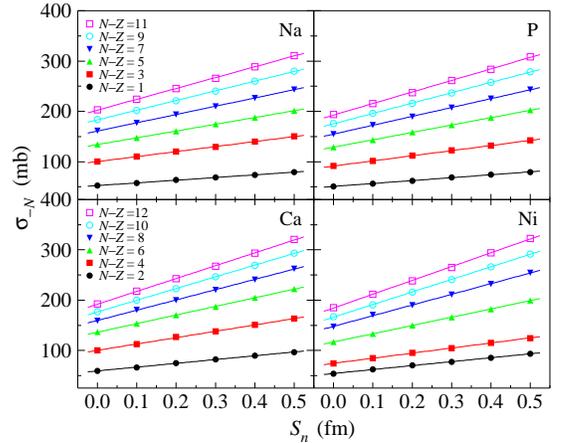}
\caption{(Color online). Correlation between $\sigma_{-N}$ and $S_{n}$
for Na, P, Ca and Ni isotopes with different $N-Z$ on $^{12}$C
target at 1{\it A} GeV calculated by the SAA model. 
The straight lines just guide the eye.}
\label{fig1}
\end{figure}
The production cross section for a specific fragment can be
calculated from
\begin{equation}\label{sfrag}
\sigma(\Delta N,\Delta Z)= \int d^{2}{\it b}P(\Delta N,
b)P(\Delta Z, b),
\end{equation}
where $\Delta N$, $\Delta Z$ are the number of abraded neutrons and 
protons from the projectile. $P(\Delta N, b)$ and $P(\Delta Z, b)$ 
are the probability distributions for the abraded neutrons and 
protons at a given impact parameter $b$, respectively. These
probability distributions are determined by a superposition of 
different binomial distributions.

The second stage is evaporation process in which the system
reorganizes due to excitation. It deexcites and
thermalizes by emission of neutron, proton and  
other light particles using
the statistical model~\cite{gaim91}.
The excitation energy for projectile spectator is estimated by a
simple relation of $E^* = 13.3 \langle \Delta A(b) \rangle$ MeV
where 13.3 is a mean $E^*$ for an abraded nucleon
from the projectile. This $E^*$ was given
by the statistical hole-energy model as described in Ref.~\cite{gaim91}.
After the evaporation stage, we can obtain the final fragments
which are comparable to the experimental data.

For the density distributions of the projectile and
target, harmonic-oscillator (HO) distributions are suitable 
for light nuclei but it is more reasonable to use the Fermi
distribution for heavier mass system~\cite{ozaw01}. 
In our calculations, the Fermi distributions are used for neutron 
and proton, respectively. They have the following form 
\begin{equation}\label{mdis}
\rho_{i}(r)=\frac{\rho_{i}^{0}}{1+
\exp(\frac{r-C_{i}}{t_{i}/4.4})},
i=\mbox{n},\mbox{p},
\end{equation}
where $\rho_{i}^{0}$ is the normalization constant which ensures
that the integration of the density distribution equals the numbers
of neutron ($i$=n) and proton ($i$=p); $t_{i}$ is the diffuseness
parameter; $C_{i}$ is the half the density radii and determined by
droplet model~\cite{myer83}
\begin{equation}\label{ci}
    C_{i}=R_{i}[1-(b_{i}/R_{i})^2], i=\mbox{n},\mbox{p},
\end{equation}
with $b_{i}=0.413t_{i}$, $R_{i}$ being the equivalent sharp
surface radii of neutron and proton. $R_{i}$ and $t_{i}$ are given 
by the droplet model. 

Neutron removal cross section ($\sigma_{-N}$) is defined as the
summation of production cross sections for fragments with the same 
proton number as the projectile, but less neutron number than the
projectile, i.e. collision processes with one or more neutrons 
being removed from the projectile:
\begin{equation}\label{nabr}
    \sigma_{-N}= \sum_{i=1}^{N}\sigma(i,0),
\end{equation}
where $N$ is the neutron number of the projectile.  
$\sigma(i,0)$ is the fragment production cross section 
with $\Delta N=i$ and $\Delta Z=0$ given by Eq.~(\ref{sfrag}). 
It should be pointed out that $\sigma(i,0)$ decreases very quickly as 
$i$ increases. The value of $\sigma(N,0)$ will be very close to zero 
for most projectiles. Since the charge number of the fragments equals 
to that of the projectile, it indicates that peripheral reactions 
will be the main contributions of $\sigma_{-N}$.
Thus $\sigma_{-N}$ maybe sensitive to the neutron distribution in
the nuclear surface. Since the reaction cross section
reflects the information about the overall size of a nucleus, 
$S_{n}$ of Na isotope determined by the 
interaction cross section ($\sigma_I$) has an error in the order of 
$\pm0.06 \sim$ $\pm0.3$fm when the error of $\sigma_I$ 
is around 1$\sim$5\%~\cite{suzu95}. 
Below we will demonstrate that determining $S_{n}$ through 
$\sigma_{-N}$ measurement may reduce the uncertainty.

In Ref.~\cite{trzc01}, Trzci\'{n}ska {\it et al} found that the half
density radii for neutrons and protons in heavy nuclei were
almost the same, but the diffuseness parameter for neutrons was
larger than that for protons. 
To study the dependence between $\sigma_{-N}$ and $S_{n}$,
the neutron diffuseness parameter in the droplet model is changed
to obtain different value of $S_{n}$ for a nucleus.
Six values of $S_{n}$ varying from 0.0 fm to 0.5 fm with step 
of 0.1 fm are chosen.
Different values of $\sigma_{-N}$ are obtained in the SAA
calculation with different $S_{n}$ of the projectile.
Thus we can obtain the correlation between $\sigma_{-N}$ and
$S_{n}$.

\begin{figure}[b]
\centering
\includegraphics[width=7.2cm,angle=0.]{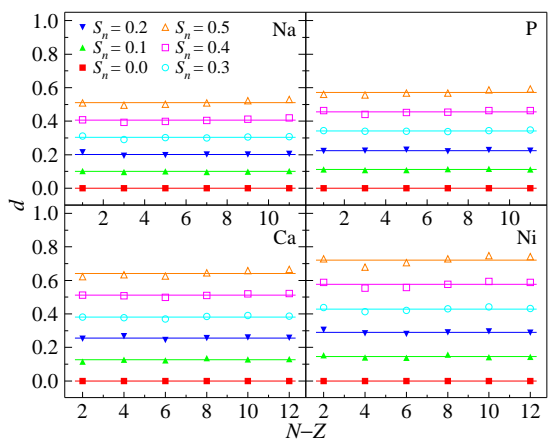}
\caption{(Color online). The $N-Z$ dependence of the difference 
factor $d$ for Na, P, Ca and Ni isotopes with $S_{n}$ varying 
from 0.0 fm to 0.5 fm. 
The horizontal lines just guide the eye.} 
\label{fig2}
\end{figure}

In this work, calculations have been done for
$^{23,25,27,29,31,33}$Na, $^{31,33,35,37,39,41}$P,
$^{42,44,46,48,50,52}$Ca and $^{58,60,62,64,66,68}$Ni projectiles 
on $^{12}$C target at 1$\it A$ GeV. The results are shown in
Fig.~\ref{fig1}. When $S_{n}$ varies from 0.0 fm to 0.5 fm, 
we can see that $\sigma_{-N}$ is increasing linearly for all 
the isotopes calculated. 
As its definition shows, $S_{n}$ is determined by
the radii of protons and neutrons. Since most nuclei have precise
proton radii, the value of neutron removal cross section for 
$S_{n}=0$ or $r_{n}=r_{p}$ ($\sigma^0_{-N}$) could be calculated 
by the SAA model using the experimental $r_p$ data .
To study the relative increase of $\sigma_{-N}$ with respect 
to $\sigma^0_{-N}$, we define a difference factor 
\begin{equation}
d=\frac{\sigma_{-N}-\sigma^0_{-N}}{\sigma^0_{-N}},
\end{equation}

\begin{figure}[t]
\includegraphics[width=7.2cm,angle=0.]{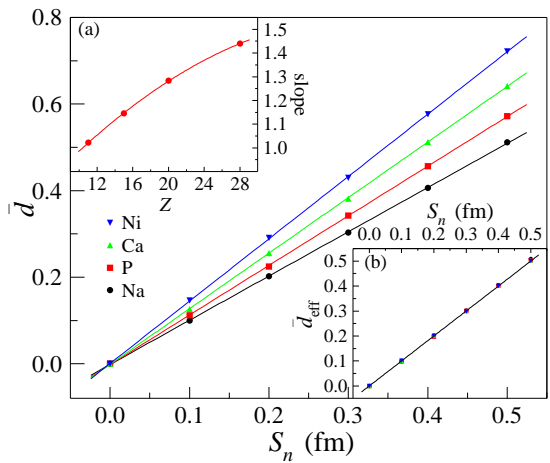}
\caption{(Color online). 
The dependece of $\bar{d}$ with $S_{n}$ for Na, P, Ca and Ni 
isotopes. The lines are the linear fit results.
Inset (a) is the dependence of the slope of $\bar{d}$ with $Z$.
The line is the fitted function $f(Z)$.
Inset (b) is the correlation between $\bar{d}_{eff}$ and $S_{n}$ for 
Na, P Ca and Ni isotopes. The line is the function of $y=x$. 
For details see the text.}
\label{fig3}
\end{figure}

The systematic behavior of $d$ with respect to $S_{n}$
for all the isotopes are calculated and shown in Fig.~\ref{fig2}.
From this figure, we found that the values of $d$ for one isotope
is almost constant as the horizontal lines showing, especially for 
$S_{n}<0.4$ fm. For one isotope with same neutron skin thickness, 
the mean uncertainty of $d$ is less than $\pm1.2\%$ for $S_{n}<0.4$ fm.
It indicates that the relative increases of $\sigma_{-N}$ 
with respect to $\sigma^0_{-N}$ are almost same for one isotope having 
different neutron excess ($N-Z$) when the neutron skin thickness 
changes from 0 fm to 0.4 fm.
The mean values of $d$ for one isotope increase as $S_{n}$ increases.
We denote this mean value of $d$ by $\bar{d}$. The dependence of 
$\bar{d}$ on $S_{n}$ for the four isotopes are shown in Fig. 3.
Clear linear relationship between $\bar{d}$ and $S_{n}$ can be seen.
We fitted the calculated results by using a function of
$\bar{d}=slope\times S_{n}$. The dependence of the obtained slope 
parameters on the charge number of the isotope is plotted in the 
inset (a) of Fig.~\ref{fig3}.
A second order polynomial function $f(Z)$ was used to fit 
the results. Then we define an effective difference factor
\begin{eqnarray}
\bar{d}_{eff}=\bar{d}/f(Z). 
\end{eqnarray}
with $f(Z)=0.5848+0.04554Z-5.356\times10^{-4}Z^2$ 
being the fitted function. The reduced $\chi^2$ 
of the fit is $7.6\times10^{-6}$.
From the dependence of $\bar{d}_{eff}$ on $S_{n}$ as 
shown in the inset (b) of Fig.~\ref{fig3}, $\bar{d}_{eff}$ always
equals the neutron skin thickness for all the nuclei.
If we could extract $\bar{d}_{eff}$ from experimental 
measurements, then the neutron skin thickness could be determined.
To calculate $\bar{d}_{eff}$, $\sigma_{-N}$ and $\sigma^0_{-N}$
are necessary. $\sigma^0_{-N}$ can be determined by the SAA model 
if we have the experimental proton radii data. Thus if $\sigma_{-N}$ 
is measured experimentally, we could obtain $\bar{d}_{eff}$. 
This indicates that $S_{n}$ can be extracted from the 
experimental measurement of $\sigma_{-N}$. The accuracy of the extracted 
$S_{n}$ is mainly determined by the uncertainty of $\sigma_{-N}$. 
Estimation based on Fig.~\ref{fig3} show that 5$\%$ error of $\sigma_{-N}$ 
corresponds to around 0.05 fm error in $S_{n}$ for Na isotopes. 
If the error of $\sigma_{-N}$ could be smaller, then the uncertainty 
of the deduced $S_{n}$ could become smaller also.

Based on above discussion, a general phenomenon is shown 
that $\sigma_{-N}$ and $S_{n}$ have strong linear correlation 
for neutron rich nuclei. Owing to the small uncertainty of 
$S_{n}$ extracted from $\sigma_{-N}$ data, 
$\sigma_{-N}$ could be used as a 
observable to determine $S_{n}$. So we propose to use
$\bar{d}_{eff}$ as a new observable to extract $S_{n}$
experimentally. 
But it should be pointed out that Pygmy Dipole Resonance (PDR)
exists in neutron rich nuclei, which has been interpreted as
a resonant oscillation of neutron skin against the remaining
isospin saturated neutron-proton core\cite{leis01,piek06,vret08}. 
In this excitation mode, the Fermi distribution could not describe 
the nuclear densities of neutron rich nuclei. 
We are focusing on the relationship between $S_{n}$
and $\sigma_{-N}$, similar correlation will also exist 
even if the shape of density distribution is different 
since the RMS radii of neutron and proton are 
quantities describing the averaged distribution radii of the nucleon. 
Further investigations on the effect of the density shape over
$\sigma_{-N}$ and the uncertainty of the deduced $S_{n}$ should be done.

In summary, we have studied the relation between neutron removal 
cross section and neutron skin for neutron 
rich nuclei using SAA model. 
Strong linear correlation between 
$\sigma_{-N}$ and the neutron skin thickness for neutron rich 
nuclei have been observed. Thus measuring $\sigma_{-N}$ will 
be a very good method for studying the neutron skin size in nuclear
surface. 
By defining an effective difference factor, 
further analysis suggests that the neutron skin thickness could 
be extracted from the experimental measurement of $\sigma_{-N}$.
We find that the uncertainty of $S_{n}$ deduced from $\sigma_{-N}$ 
data is quite small. So we propose to use 
$\bar{d}_{eff}$ as a new quantitative measure 
for neutron skin thickness in neutron rich nuclei.

This work is supported by National Natural Science Foundation 
of China under contract No.s 10775168, 10775167, 10979074 and 10975174,
Major State Basic Research Development Program in China under 
contract No. 2007CB815004, Knowledge Innovation Project of 
Chinese Academy of Sciences under contract No. KJCX3.SYW.N2,
and the Shanghai Development Foundation for Science and 
Technology under contract No. 09JC1416800.


\footnotesize
\begin{thebibliography}{}
\bibitem{brow00} B. A. Brown, Phys. Rev. Lett {\bf 85}, 5296 (2000).

\bibitem{furn02} R. J. Furnstahl, Nucl. Phys. {\bf A706}, 85 (2002).

\bibitem{yosh04} S. Yoshida and H. Sagawa, Phys. Rev. C {\bf 69}, 024318 (2004).

\bibitem{dani03} P. Danielewicz, Nucl. Phys. {\bf A727}, 233 (2003).

\bibitem{chen05} L. W. Chen, C. M. Ko, and B. A. Li, Phys. Rev. C {\bf 72}, 064309 (2005).

\bibitem{cent09} M. Centelles {\it et al}., Phys. Rev. Lett. {\bf 102}, 122502 (2009).

\bibitem{horw01} C. J. Horowitz and J. Piekarewicz, Phys. Rev. Lett {\bf 86}, 5647 (2001).

\bibitem{poll99} S. J. Pollock and M. C. Welliver, Phys. Lett. B {\bf 464}, 177 (1999).

\bibitem{fric95} G. Fricke {\it et al}., At. Data Nucl. Data Tables {\bf 60}, 177 (1995).

\bibitem{horo01} C. J. Horowitz {\it et al}., Phys. Rev. C {\bf 63}, 025501 (2001).

\bibitem{kara02} S. Karataglidis {\it et al}., Phys. Rev. C {\bf 65}, 044306 (2002).

\bibitem{clar03} B. C. Clark, L. J. Kerr, and S. Hama, Phys. Rev. C {\bf 67}, 054605 (2003).

\bibitem{brow07} B. A. Brown {\it et al}., Phys. Rev. C {\bf 76}, 034305 (2007).

\bibitem{trzc01} A. Trzci\'{n}ska  {\it et al}., Phys. Rev. Lett. {\bf 87}, 082501 (2001).

\bibitem{suzu95} T. Suzuki {\it et al}., Phys. Rev. Lett. {\bf 75}, 3241 (1995).

\bibitem{mach08} C. W. Ma {\it et al}., Chin. Phys. B {\bf 17}, 1216 (2008).

\bibitem{broh94} T. Brohm and H. H. Schmidt, Nucl. Phys. {\bf A569}, 821 (1994).

\bibitem{blan95} B. Blank {\it et al}., Nucl. Phys. {\bf A588}, 171c (1995).

\bibitem{jong98} M. de Jong {\it et al}., Nucl. Phys. {\bf A628}, 479 (1998).
 
\bibitem{fang01} D. Q. Fang {\it et al}., Eur. Phys. J. A {\bf 10}, 381 (2001).

\bibitem{fang07} D. Q. Fang {\it et al}., J. Phys. G. {\bf 34}, 2173 (2007).

\bibitem{gaim91} J. J. Gaimard and K. H. Schmidt, Nucl. Phys. {\bf A531}, 709 (1991).

\bibitem{ozaw01} A. Ozawa, T. Suzuki, and I. Tanihata, Nucl. Phys. {\bf A693}, 32 (2001).

\bibitem{myer83} W. D. Myers and K. H. Schmidt, Nucl. Phys. {\bf A410}, 61 (1983).

\bibitem{leis01} A. Leistenschneider {\it et al}., Phys. Rev. Lett. {\bf 86}, 5442 (2001).

\bibitem{piek06} J. Piekarewicz, Phys. Rev. C {\bf 73}, 044325 (2006).

\bibitem{vret08} D. Vretenar, N. Paar, T. Marketin and P. Ring, J. Phys. G. {\bf 35}, 014039 (2008).

\end{thebibliography}
\end{document}